\documentstyle[aps,pre,epsfig]{revtex}

\begin{document}
\draft

\title{{\bf Second virial coefficient for the Landau diamagnetism of a two component plasma}}
\author{M.Steinberg, W. Ebeling, and J. Ortner}


\date{\today}
\maketitle

\begin{abstract}
This paper investigates  the density expansion of the thermodynamic properties of a two component plasma under the influence of a weak constant uniform magnetic field. We start with the fugacity expansion for the Helmholtz free energy. The leading terms with respect to the density are calculated by a perturbation expansion with respect to the magnetic field. We find a new magnetic virial function for a low density plasma which is exact in quadratic order with respect to the magnetic field. Using these results we compute the magnetization and the magnetic susceptibility.

\end{abstract}

\pacs{52.25.Kn, 05.30.-d, 05.70.Ce, 71.45.Gm}

\section{Introduction}
The topic of matter in magnetic fields has received much attention. The magnetic field has a great effect on the individual motion of charged particles. In classical mechanics the motion of the free particles in a magnetic field can be described by circular orbits. The frequency associated with the rotation is the cyclotron frequency $\omega_c=eB_0/m_e$. In quantum mechanics the motion perpendicular to the magnetic field is quantized with the corresponding energy eigenvalues $E_{\perp}=\hbar\omega_c(n+1/2)$ \cite{Landau}.  A wide range of subsequent investigations covers the properties of atoms and molecules in magnetic fields. Their study is motivated by the astrophysical implications concerned with the physics of pulsars and neutron stars, but also by its application in quantum chaos \cite{ScSc98,RuWu94,FrWi89}.
The calculation of the energy spectrum of a hydrogen atom in strong magnetic fields has been tackled by various authors \cite{Wunner1,Potekhin,Liberman&Johansson}. One of the complications found in the theory is the coupling of the center of mass motion with the relative motion \cite{Potekhin}. This also complicates the calculation of the thermodynamic functions of a low density plasma. However, this effect becomes important at strong fields only. Throughout this paper we will consider the weak-field limit in which case this effect becomes negligible. In this limit the proton mass is considered to be infinite. 

Although the magnetic field affects the individual motion of the particles, there is no influence of the magnetic field on the equilibrium properties of a classical charged particle system. This follows from the Bohr-van-Leeuwen theorem \cite{Bohr,vanLeeuwen}, which can be easily derived by changing the variable in the momentum integrals in such a way that one works with the variable ${\bf \pi}={\bf p}-e{\bf A}$. As a result of this all equilibrium properties are independent of the magnetic field. In quantum mechanics this argument is no longer valid, since the momentum operator  and the coordinate operator of a particle do not commute. A common example of an equilibrium value which depends on the magnetic field is the magnetization of an electron gas. It was shown by Landau \cite{Landau2} that the low field magnetization of a spinless electron gas, in Boltzmann statistics, is
\begin{equation}
M_{\rm orb}=-\frac{n e \beta \hbar^2 \omega_c}{12m} \;,
\end{equation}
i.e., the so called orbital part contributes to a diamagnetic response. However, the full response of the free (noninteracting) electron gas, including the spin part, is paramagnetic. 

The magnetization of a system of charged particles is a boundary effect. In classical mechanics the magnetization induced by the motion of the bulk electrons is cancelled by the magnetization connected with the surface current. Again in quantum mechanics this statement is not valid anymore. Landau used a perturbation expansion of the free energy with respect to the magnetic field, to circumvent difficulties due to the boundary effects. In doing so the electrons at the boundary of the system were neglected. Then the magnetization is found as the derivative of the free energy with respect to the magnetic field. Another approach has been chosen by Teller \cite{Te31}. He calculated the current at the boundary of the system, produced by the motion of the elctrons under the influence of uniform magnetic field. From this he computed the magnetization of the system and was able to show the equivalence of his method and Landau's approach. However, as Teller already pointed out Landau's method is much better suited for more complicated problems.  

The difficulties connected with the boundary effects are perhaps one of the reasons for the few results concerning the equilibrium statistical mechanics of a low density quantum plasmas embedded in an external magnetic field. A first attempt beyond Landau has been pursued by Alastuey and Jancovici. They studied, by means of a Wigner-Kirkwood expansion, the magnetic properties of a nearly classical one component plasma (OCP) in two and three dimensions in the weak field \cite{AlJa79} as well as in the strong field limit \cite{AlJa80}. Related problems were treated by Cornu \cite{Cornu} and Boose and Perez \cite{Boose&Perez} who derived a formally exact virial expansion of the EOS of a multicomponent plasma by using the Feynman Kac path-integral representation of the grand-canonical ensemble.

This paper is aimed to calculate the magnetic properties of a quantum plasma in the low density limit. These systems are characterized by a small coupling parameter $\Gamma$, which is given by
\begin{equation}
\Gamma =\frac{e^2}{4 \pi \epsilon_0 kT d} \, ,
\end{equation}
where $d=(3 / 4\pi n)^{1/3}$ is the mean distance between the particles. We follow the method of Landau for the calculation of the magnetization. The starting point is the fugacity expansion of the Helmholtz free energy. In a previous work \cite{StOrEb98} the authors have performed a perturbation expansion for the equation of state of a low density plasma up to the order $n^2 e^4$, which is valid at arbitrary magnetic field strength. In the present paper we perform an expansion of the thermodynamic functions for a low density plasma ($\Gamma<1$) up to the order $n^2$ and calculate the coefficients of this expansion in quadratic order with respect to the magnetic field, without making any approximation with respect to the Coulomb problem. Using a diagrammatic language this can be restated as the calculation of all ladder diagrams expanded to second order in the magnetic field. In doing so we will consider the ${\bf B}$ and ${\bf B}^2$ terms of the Hamiltonion separately. We will see that the separate contributions are divergent, only the sum of all contributions give a convergent expression. This is the price which we have to pay within the present method circumventing the calculation of boundary effects. \\

This paper is organized as follows. In section II, we discuss the representation of thermodynamic functions by a fugacity expansion. The second virial coefficient of a magnetized plasma will be discussed in section III. In the first part of section III, we present in more detail the calculation of the electron-ion contribution to the thermodynamic functions in the case of an infinite proton mass and in the second part an analytical continuation will lead us to the electron-electron contribution. In the third part of section III the asymptotic behavior of the new proposed magnetic virial functions will be studied. Finally, the derived results are used to compute the magnetization and the magnetic susceptibility in section IV.

\section{Representation of the Thermodynamic Functions by a Fugacity Expansion}

In this section, we briefly present the general method used in this work and give the exact results derived in an earlier work \cite{StOrEb98}. Let us consider a two-component charge-symmetrical system of N spin half particles of charge (-e) and mass $m_e$ and N spin half particles of charge e and mass $m_i$. The Hamiltonoperator of our system consists of two particle contributions. Each pair of species $a$ and $b$ contributes
\begin{eqnarray}
\label{nd75} H_{ab}^\lambda = \left( \frac{({\bf p}_a-e_a{\bf A}_a)^2}{2m_a}+\mu_{B}^a B_0 \sigma_z \right) + \left( \frac{({\bf p}_b-e_b{\bf A}_b)^2}{2m_b}+\mu_{B}^b B_0 \sigma_z \right) +\lambda V_{ab}({\bf r}) \, , \nonumber\\
\hspace{1cm} \sigma_z=-1,+1
\end{eqnarray} 
with the Coulombic interaction potential 
\begin{equation}
V_{ab}({\bf r}) = \frac{e_a e_b}{4 \pi \epsilon_0 r} \, .
\end{equation}
Here $ H_{ab}$ is the Hamiltonoperator of the two particle system and $H_{ab}^0$ of the noninteracting system. The additive term $\mu_{B}^a B_0 \sigma_z $ takes into account the coupling between the intrinsic magnetic moment ($\mu_{B}^a = e_a \hbar /(2m_a)$) of the charged particles and the magnetic field. \\
We suppose that the pressure can be split into ideal and interaction contributions 
\begin{equation}
\label{2.0} p=p_{id} + p_{int} \, .
\end{equation}
In the case without Coulomb interaction $e^2 = 0$ the pressure and the particle density of the  plasma in a homogeneous magnetic field ${\bf B}=(0,0,B_0)$ are given by a sum of Fermi integrals over all Landau levels n 
\begin{equation}
\label{2.1} p_{id} = kT \sum_a \, \frac{2x_a}{\Lambda_a^3} \, {\sum_{n=0}} ^\prime f_{\frac{1}{2}}(\ln \left(z_n^a\right)) \, \, , \hspace{1cm} n=\sum_{a} \frac{2x_a}{\Lambda_a^3} \, {\sum_{n=0}} ^\prime \, f_{-\frac{1}{2}}(\ln \left( z_n^a \right))
\end{equation}
($x_a=\hbar \omega_c^a/(2kT)$ with $\omega_c^a=|e_a|B_0/m_a$, $\Lambda_a =h / \sqrt{2\pi m_akT}$, and $z_n^a=\exp{[\beta(\mu-n\hbar\omega_c^a)]}$). The prime indicates the double summation due to the spin degeneracy except for the $n=0$ level. \par
The interaction part of the pressure will be expressed in terms of a fugacity expansion which will be truncated after the second virial coefficient \cite{Vedenov&Larkin,Larkin,BaEb71,EKK,Eb68}
\begin{equation}
\label{nd8} \beta  p_{int} = \frac{\kappa^3}{12 \pi} + \sum_{ab} {\tilde z}_a {\tilde z}_b \, \left(  \frac{\pi}{3} \lambda^3_{ab} \xi^3_{ab} \ln(\kappa \lambda_{ab}) + \frac{\pi}{2} \beta^3 \frac{e_a^2}{4\pi \epsilon_0} \frac{e_b^4}{(4\pi \epsilon_0)^2}+ B_{ab} \right) +  0({\tilde z}^{5/2} \ln {\tilde z}\,) ,
\end{equation} 
where we have introduced the modified fugacities
\begin{equation}
\label{nd6} {\tilde z}_a = z_a \, \frac{2}{\Lambda_a^3} \, \frac{x_a}{\tanh(x_a)} \, ,
\end{equation}
in order to have ${\tilde z}_a \rightarrow n_a$ in the limit of small densities. The first term on the r.h.s of Eq.(\ref{nd8}) is the Debye contribution in the grand canonical ensemble. The squared inverse Debye radius is given by $\kappa^2=\beta(e^2/\epsilon_0) ({\tilde z}_e + {\tilde z}_i)$. Since it is a classical contribution the Debye term does not depend on the field. In the limit of small densities Eq.(\ref{nd8}) coincides with the formally exact virial equation of state derived by Cornu \cite{Cornu} and Boose and Perez \cite{Boose&Perez}. We now try to extend these calculations and focus on the calculation of the second virial coefficient $B_{ab}$. In order to avoid convergence problems let us in a first approach cut the Coulomb tail at large distances, i.e. $V_{ab}(r)=0$ if $r>R$. Then the second virial coefficient reads
\begin{equation}
\label{nd6.44}  B_{ab} = \frac{1}{2 \Omega} \left( \frac{\Lambda_a^3}{2} \frac{\tanh(x_a)}{x_a} \right)   \left( \frac{\Lambda_b^3}{2} \frac{\tanh(x_b)}{x_b} \right)    \, {\bf Tr} \, (e^{-\beta \widehat H_{ab}} -   e^{-\beta \widehat H_{ab}^{0}}) \, .
\end{equation} 
This function will be studied in more detail in the next section. 
\section{Second virial coefficient of a magnetized plasma}
It is convenient to divide the second virial coefficient into the direct part $B_{ab}^d$ and the exchange part $B_{aa}^{ex}$
\begin{equation}
B_{ab} =B_{ab}^d+B_{aa}^{ex} \delta_{ab}\, ,
\end{equation} 
and to compute them separately. The exchange part of $B_{ab}$ as defined by Eqs.(\ref{nd6.44}) with the Hamiltonoperator given by Eq.(\ref{nd75}) is convergent while if $R\rightarrow \infty$ the direct part is divergent. This is due to the long range behavior of the Coulomb interaction which leads to collective effects. In order to include these collective effects one has to perform a screening procedure, which may then lead to convergent expression for $B_{ab}$. Such a technique is well established in the zero magnetic field case \cite{Vedenov&Larkin,Larkin,BaEb71,EKK,Eb68} and can be easily extended to the nonzero magnetic field case. In general, $B_{ab}$ is an analytic function of the interaction parameter $\xi_{ab}$ \cite{Eb68}, defined by
\begin{equation}
\xi_{ab}=-\frac{e_a e_b}{4\pi \epsilon_0 k T \lambda_{ab}} \, ,
\end{equation}
with $\lambda_{ab}=\hbar/\sqrt{2m_{ab}kT}$ and $m_{ab}=m_a m_b/(m_a+m_b)$ beeing the effective mass. Hence $B_{ab}$ may be written as a Taylor expansion. Using the methods as described in \cite{Vedenov&Larkin,Larkin,BaEb71,EKK,Eb68} we derived in our earlier work \cite{StOrEb98} the lowest order contribution to $B_{ab}$. As in the zero field case we write the direct part of the second virial coefficient of the plasma in the following form
\begin{equation}
B_{ab}^d = B'_{ab} + B''_{ab} \, ,
\end{equation}
where the contributions of second and third order in $\xi_{ab}$ are included in $B'_{ab}$, with 
\begin{equation}
B'_{ab} = - \frac{1}{8} \pi^{3/2} \lambda_{ab}^3 \xi_{ab}^2 h_2(x_a, x_b) 
- \frac{\pi}{3} \left(\frac{C}{2} + \log3 - \frac{1}{2}\right) \lambda_{ab}^3 \xi_{ab}^3 h_3(x_a,x_b) \, .
\end{equation}
In general the magnetic field correction $h_{2,3}$ satisfies $h_{2,3} = 1$ if the magnetic field $B = 0$. The second order term has been found in \cite{StOrEb98} and is explicitely given by
\begin{equation}
\label{mwt9}  h_2(x_a,x_b) = \left( \frac{1}{2} + \frac{4}{\pi} \int_0^1 dt \, \sqrt{t(1-t)} \, (y_a+y_b) \, \frac{{\rm arctanh}{\sqrt{1-(y_a+y_b)}}}{\sqrt{1-(y_a+y_b)}}\right) \, ,
\end{equation}
with $ y_{a,b} = \lambda_{aa,bb}^2 \, \sinh(x_{a,b}t) \, \sinh(x_{a,b}(1-t))/(\lambda_{ab}^2 \, t(1-t) \, 2 x_{a,b} \, \sinh(x_{a,b})) $. The magnetic field correction $h_3$ is so far not exactly known. In the limit of zero field $h_3=1$ holds; furthermore, in the next section we will derive an expression for $h_3$ in the weak field limit. \\

Formally the higher order contributions may be expressed by a resolvent expansion \cite{EKK}

\begin{eqnarray}
B''_{ab} & = & \frac{1}{2 \Omega} \left( \frac{\Lambda_a^3}{2} \frac{\tanh(x_a)}{x_a} \right)   \left( \frac{\Lambda_b^3}{2} \frac{\tanh(x_b)}{x_b} \right) P''   \nonumber\\
& & \, {\bf Tr} \,  \sum_k \frac{1}{2\pi i} \int_C dz \, e^{-\beta z} \left[\frac{1}{H_{ab}^0 - z} V_{ab}\right]^k \frac{1}{H_{ab}^0-z} \, .
\end{eqnarray}
The contour integral may be taken in the sense of an inverse Laplace transform. The operator $P''$ means that all terms of order less than $\xi^4$ have to be omitted, since they have already been taken into account in $B'_{ab}$. The series may then be written in the general form 
\begin{equation}
B''_{ab} = 2 \pi^{3/2} \lambda_{ab}^3 \sum_{k=4}^{\infty} 
\frac{ \zeta(k-2) h_k(x_a,x_b)}{\Gamma(1+k/2)} \left(\frac{\xi_{ab}}{2}\right)^k \, .
\end{equation}
The functions $h_k$ expressing the magnetic corrections satisfy the zero field condition
\begin{equation}
h_k(0,0) = 1 \, .
\end{equation}
Therefore in the zero magnetic field case an exact calculation of the convergent second virial coefficient is possible in agreement with earlier work \cite{EKK,Eb68}. 

An alternative expression for the field free virial coefficient which we may refer to as $B^0_{ab}$ may be obtained by introducing the quantum virial function $Q^0 (\xi_{ab})$ \cite{Eb68} according to
\begin{equation} 
B^0_{ab}=2 \pi \lambda_{ab}^3 Q^0 (\xi_{ab})\, , 
\end{equation}
with
\begin{equation}
Q^0(\xi_{ab}) = -\frac{1}{8} \sqrt \pi \xi^2_{ab} -\frac{1}{6} \xi_{ab}^3 \left(\frac{{\bf C}}{2} +\ln 3 -\frac{1}{2}\right)+ \sqrt \pi \sum_{k=4}^{\infty} \frac{ \zeta(k-2) }{\Gamma(1+k/2)} \left(\frac{\xi_{ab}}{2}\right)^k \, .
\end{equation}
Note that the second order term may be included into the series, since $\zeta(0)=-1/2$. \par

Now let us discuss the exchange part. Again, as it was shown in Ref.\cite{StOrEb98} this contribution may be written in  a Taylor expansion
\begin{equation}
B_{aa}^{ex}  = \pi^{3/2} \lambda_{aa}^3 \sum_{k=0}^{\infty} \left(1-2^{2-k} \right) \frac{ \zeta(k-1) b_{k}(x_a)}{\Gamma(1+k/2)} \left(\frac{\xi_{aa}}{2}\right)^k \, .
\end{equation}  
Here we have included the terms with nonpositive arguments in the $\zeta$ function. In particular we have used the relation 
\begin{equation}
\lim_{k\rightarrow 2} (1-2^{2-k}) \zeta(k-1)=\ln 2  \, . 
\end{equation}
The zero field results are reproduced, since we have $b_k(0)=1$  and they may be written, after introducing the exchange virial function $E^0(\xi_{aa})$, as 
\begin{equation} 
B_{aa}^{0 ex}= - \pi \lambda^3_{aa} E^0(\xi_{aa}) \, , 
\end{equation}
with
\begin{equation}
E^0(\xi_{aa}) = \sqrt \pi \sum_{k=0}^{\infty} \left(1-2^{2-k} \right) \frac{ \zeta(k-1) }{\Gamma(1+k/2)} \left(\frac{\xi_{aa}}{2}\right)^k \, .
\end{equation}
The influence of the magnetic field on the exchange part has been studied in \cite{StOrEb98} for the lower order terms $b_0$, $b_1$, and $b_2$. The following analytical expression were derived
\begin{equation}
\label{2001} b_0 (x_a) = \frac{\cosh(2x_a)}{\cosh^2(x_a)} \frac{\tanh(x_a)}{x_a} \, ,
\end{equation} 
and
\begin{equation}
\label{2002} b_1 (x_a) = \frac{\cosh(2x_a)}{\cosh^2(x_a)} \frac{\tanh(x_a)}{x_a} \frac{{\rm arctanh} \sqrt{ 1-\frac{\tanh(x_a)}{x_a}}}{\sqrt{1-\frac{\tanh(x_a)}{x_a}}} \, .
\end{equation} 
For an integral representation of $b_2$ we refer the reader to Eq.(C4) of \cite{StOrEb98}. 

\subsection{Expansion in the weak field limit for the ion-electron interaction}
We shall find an expansion of the second virial coefficient in the weak-field limit. The magnetic field is now assumed to be a small perturbation to the field free Coulomb problem. In this case we can use the already established results for the second virial coefficient in the absence of a field \cite{Eb68}. 
Due to the invariance of the thermodynamic functions under the transformation ${\bf B} \rightarrow -{\bf B}$ the first correction to the field-free results will be quadratic in the magnetic field. This can also be verified in the ideal contribution (Pauli spin magnetism and Landau diamagnetism). \\
Let us consider a hydrogen plasma with an infinite proton mass. This is a reasonable approximation in the weak field limit, as the proton frequency is proportional to the inverse mass of the proton. We chose the symmetric gauge ${\bf A} = 1/2 \, \left( {\bf B} \times {\bf r} \right)$. Then the Hamiltonian in relativ and center of mass coordinates takes the form
\begin{equation}
H_{ei} = \frac{{{\bf P}}^2}{2m_i}+ \frac{{{\bf p}}^2}{2m_{e}}+\frac{m_e \omega_c^2}{8} \rho^2-i\frac{\hbar \omega_c}{2} \frac{\partial}{\partial \phi}+\mu_{B}^e B_0 \sigma_z - \frac{e^2}{4\pi \epsilon_0 r}\, , \nonumber\\
\end{equation} 
where $\omega_c$ is the electron cyclotron frequency. The elctron-ion contribution $B_{ei}^d$ to the second virial coefficient is given by the following trace 
\begin{equation}
B_{ei}^d = \, \frac{1}{2 \Omega} {\bf P}^{k^\prime} \left( \frac{\Lambda_i^3}{2} \right)   \left( \frac{\Lambda_e^3}{2} \frac{\tanh(x_e)}{x_e} \right) \, {\bf Tr} \, (e^{-\beta H_{ei}} -   e^{-\beta  H_{ei}^{0}}) \, .
\end{equation}
As in the zero magnetic field case we have defined an operator ${\bf P}^{k^\prime}$ that takes into account the divergency, by omitting all terms of order $e^{k}$ with $k < k^\prime$.

The trace over the center of mass coodinates and over the spin variable is readily carried out. Again we use the resolvent representation to obtain the following contribution
\begin{equation}
B_{ei}^d = 4 \pi^{3/2} \lambda^3_{e} \frac{\sinh(x_e)}{x_e} {\bf P}^{k^\prime} \int_C \frac{dz}{2\pi i} e^{-\beta z} {\bf Tr} \left(\frac{1}{h_{ei}-z}-\frac{1}{h^0_{ei}-z} \right) \, ,
\end{equation}
with $\lambda_e=\hbar/\sqrt{2m_ekT}$. Here we have introduced the Hamiltonoperators for the free relative motion
\begin{equation}
h^0_{ei}=\frac{{{\bf p}}^2}{2m_e}+\frac{m_e \omega_c^2}{8} \rho^2-i\frac{\hbar \omega_c}{2} \frac{\partial}{\partial \phi} \, ,
\end{equation}
and for the relative motion of the interacting particles
\begin{equation}
h_{ei}=h^0_{ei}-\frac{e^2}{4\pi \epsilon_0 r} \, .
\end{equation}
We are interested in the case of a weak magnetic field without making any approximation with respect to the Coulomb problem. For that we expand $B_{ei}$ in powers of $x_e=\hbar \omega_c/2kT$. It can be easily shown that the linear term is equal to zero and the first nonvanishing term is proportional to $x_e^2$. This contribution may be written as
\begin{equation}
\label{2201} B_{ei}^d = \left( 1+\frac{x_e^2}{6}\right) B^0_{ei} + B^1_{ei}+B^2_{ei} \, .
\end{equation}
The first term comes from the expansion of the normalising constant and therefore the trace is solely given by the zero field result \cite{Eb68}. In order to take into account the infinite proton mass in $B^0_{ei}$ one has to replace $\lambda_{ei}$ by $\lambda_{e}=\hbar/\sqrt{2m_ekT}$ and, hence, $\xi_{ei}$ by $\xi_e=-e_e e_i/(4\pi \epsilon_0 kT \lambda_{e})$. The other two terms, beeing of the order $O(B^2)$, are the result of an expansion of the trace in powers of the magnetic field and read as
\begin{equation}
B^{1}_{ei} = 4 \pi^{3/2} \lambda^3_{e} {\bf P}^{k^\prime} \int_C \frac{dz}{2\pi i} e^{-\beta z} {\bf Tr} \left(\frac{1}{h^c-z} \, \left(-i\frac{\hbar \omega_c}{2} \frac{\partial}{\partial \phi} \right) \, \frac{1}{h^c-z} \, \left( -i\frac{\hbar \omega_c}{2} \frac{\partial}{\partial \phi} \right) \, \frac{1}{h^c-z} \right)
\end{equation}
and
\begin{equation}
B^{2}_{ei} = - 4 \pi^{3/2} \lambda^3_{e} {\bf P}^{k^\prime} \int_C \frac{dz}{2\pi i} e^{-\beta z} {\bf Tr} \left(\frac{1}{h^c-z} \left(\frac{m_e\omega^2_c}{8} \rho^2\right)\frac{1}{h^c-z} \right) \, .
\end{equation}
Here $h^c={\bf p}^2/2m_{e}-e^2/4\pi \epsilon_0 r$ is the Hamiltonian for the Coulomb problem for zero magnetic field. In what follows we briefly outline the steps leading to the final result for $B^{1}_{ei}$ and $B^{2}_{ei}$. For simplicity the calculations of these contributions may be carried out separately, but as will be seen below only the sum of both gives a convergent contribution. 
\subsubsection{Calculation of $B^{1}_{ei}$}
Let us first concentrate on the calculation of $B^{1}_{ei}$. The perturbation operator has spherical symmetry. Thus it is convenient to use the eigenfunctions of the Coulomb operator. With that the calculation of the matrix elements becomes trivial. As in the zero field case we can write 
\begin{eqnarray}
\label{194} B^{1}_{ei} & = & 4 \pi^{3/2} \lambda^3_{e} {\bf P}^{k^\prime} \int_C \frac{dz}{2\pi i} e^{-\beta z} {\Bigg \{} \sum_{n=1}^\infty \sum_{l=0}^{n-1} \sum_{m=-l}^{l} \frac{1}{\left(E_n-z\right)^3} \frac{\left(\hbar \omega_c\right)^2}{4} m^2 \nonumber\\
\label{195}  & & + \sum_{l=0}^{\infty} \sum_{m=-l}^{l} \int_0^\infty dk \frac{1}{\left(\frac{\hbar^2 k^2}{2m}-z\right)^3} \frac{\left(\hbar \omega_c\right)^2}{4} m^2 \, \frac{1}{\pi} \frac{d \delta_l(k)}{d k} {\Bigg \}} \, .
\end{eqnarray}
Here we have made use of the relation between the density of states for the continuum states and the scattering phase shifts $\delta_l(k)$ of the Coulomb system. The eigenvalues of the Coulomb system read as $E_n=-1/2n^2$ and can be expressed in terms of the parameter $\xi$ by
\begin{equation}
-\beta E_n =  \left(\frac{\xi_{e}}{2}\right)^2 \frac{1}{n^2} \, .
\end{equation} 
First we compute the discrete part of the partition function, that is given by the first term in (\ref{195}) and reads
\begin{equation}
\label{196} B^{1b}_{ei} = 4 \pi^{3/2} \lambda^3_{e} \frac{x_e^2}{\Gamma(3)} {\bf P}^{k^\prime} \,  \sum_{n=1}^\infty \sum_{l=0}^{n-1} \sum_{m=-l}^{l} m^2 e^{-\beta E_n}  \, ,
\end{equation}
where we have performed the inverse Laplace transform. The summation over m and l is trivial and one immediately finds, that
\begin{equation}
\label{197} B^{1b}_{ei} = 4 \pi^{3/2} \lambda^3_{e} \frac{x_e^2}{12} \, {\bf P}^{k^\prime} \sum_{n=1}^\infty \left( n^4-n^2 \right) \exp{\left(\frac{\xi_{e}}{2n}\right)^2 } \, .
\end{equation}
By expanding the exponential and using the representation of the $\zeta$-function we obtain 
\begin{equation}
\label{198} B^{1b}_{ei} =4 \pi^{3/2} \lambda^3_{e} \frac{x_e^2}{12} {\bf P}^{k^\prime} \sum_{k=2,4,\cdots} \frac{\zeta(k-4)-\zeta(k-2)}{\Gamma(1+k/2)} \left(\frac{\xi_{e}}{2}\right)^k \, .
\end{equation}
So far we have calculated the bound state contribution to $B^{1}_{ei}$. In the next step we consider the contribution of continous spectrum. For that we need the scattering phase shifts of the field free Coulomb problem that are given by
\begin{equation}
\label{199} \frac{d}{dk} \delta_{l}(k)=-\frac{1}{k^2}\left(\frac{e^2 m_{e}}{\hbar^2} \right) \, Re \psi\left( l+1+i \left| \frac{e^2 m_{e}}{k \hbar^2} \right| \right) \, .
\end{equation}
Making use of this relation and introducing $t=\lambda^2_{e} k^2$ the second term in Eq.(\ref{195}) may be written as
\begin{equation}
\label{200} B^{1s}_{ei} =\pi^{1/2} \lambda^3_{e} x_e^2  {\bf P}^{k^\prime} \int_C \frac{dz}{2 \pi i} e^{- z} \sum_{l=0}^{\infty} \sum_{m=-l}^{l} m^2 \int_0^\infty \frac{dt}{t^{3/2}} \frac{\xi_{e}}{\left(t- z\right)^3} Re \psi\left( l+1+\frac{i}{2 \sqrt t} \left|\xi_{e} \right| \right) \, .
\end{equation}
In order to compute the sum over m and l we will expand the $\psi$-function, we have  
\begin{equation}
\label{201} Re \psi\left( l+1+\frac{i}{2 \sqrt t} \left|\xi_{e} \right| \right) = Re \sum_{k=0}^{\infty} \frac{1}{k!} \psi^{(k)}(l+1) \, \left( \frac{i}{2 \sqrt t} \left|\xi_{e} \right|\right)^k \, .
\end{equation}
Now the summation over m and l may be carried out. We obtain by introducing $\tau=l+1+s$ 
\begin{eqnarray}
\label{202} {\bf P}^{k^\prime} \sum_{l=0}^{\infty} \sum_{m=-l}^{l} m^2 \frac{1}{k!} \psi^{(k)}(l+1) & = & (-1)^k {\bf P}^{k^\prime} \sum_{l=0}^{\infty} \frac{1}{3} \left( 2 l^3+3 l^2 +l\right)  \sum_{s=0}^\infty \frac{1}{\left(l+1+s\right)^{k+1}} \nonumber\\
\label{203} & = & (-1)^k {\bf P}^{k^\prime} \sum_{\tau=1}^{\infty} \frac{1}{\tau^{k+1}} \sum_{l=0}^{\tau-1} \frac{1}{3} \left( 2 l^3+3 l^2 +l\right) \nonumber\\
\label{204} & = & (-1)^k {\bf P}^{k^\prime} \frac{1}{6} \left(\zeta(k-3)-\zeta(k-1)\right) \, .
\end{eqnarray}
Next we perform all remaining integrations. In this context we may use the following integral representation
\begin{equation}
\label{205} \int_C \frac{dz}{2\pi i} e^{-z} \int_{C^\prime} \frac{dt}{t^{3/2}} \frac{\xi_{e}}{\left(t-z\right)^3} \left( \frac{i}{2 \sqrt t} \left|\xi_{e} \right|\right)^k = \frac{2 \pi}{\Gamma((k+3)/2)} \left(\frac{\mid \xi_{e} \mid}{2} \right)^{k+1} \, .
\end{equation}
Here the contour integral $C^\prime$ in the complex $t$-plane encircles the positive real axis in the mathematical positive sense. Using Eqs.(\ref{200},\ref{204} and \ref{205}) we obtain, after shifting the summation index $k \rightarrow k-1$, the series 
\begin{equation}
\label{206} B^{1s}_{ei} =-2 \pi^{3/2} \lambda^3_{e} \frac{x_e^2}{12} \sum_{k=6} \frac{\zeta(k-4)-\zeta(k-2)}{\Gamma(1+k/2)} \left(-\frac{\mid \xi_{e} \mid}{2}\right)^k \, .
\end{equation}
Finally we sum up the bound state (\ref{198}) and the scattering state  (\ref{206}) contribution, which gives
\begin{equation}
\label{207} B^1_{ei} =2 \pi^{3/2} \lambda^3_{e} \frac{x^2}{12}  \sum_{k=6} \frac{\zeta(k-4)-\zeta(k-2)}{\Gamma(1+k/2)} \left(\frac{\mid \xi_{e} \mid }{2}\right)^k \, .
\end{equation}
Here the sum runs from $k=6$, since in this derivation the lower order terms $k<6$ would give divergent contributions . However, formally the $\zeta$ function can be extended to negative values and therefore the sum to smaller $k$ values such as $k=2,3,4$, and $5$. It will be shown below that this extension is possible and gives the exact lower order contributions. \par 
Note that the bound state contribution and the scattering state contribution  differ by a factor of 2. This general statement has been previously derived in the zero field case \cite{EKK}. It is essentially a consequence of the analyticity of the second virial coefficient $B_{ab}(\xi)$ and  expresses the fact of compensation of bound state and scattering state contributions according to Levinsons Theorem \cite{EKK}. One may also regard it as rule of obtaining scattering quantities from bound state quantities. We will employ this relation in the following section.
\subsubsection{Calculation of $B_{ei}^2$}
Again we first concentrate on the calculation of the bound state contribution. We may use the eigenfunction of the Coulomb operator to evaluate the trace. Thus we have
\begin{equation}
\label{208} B^{2b}_{ei}  =  -4 \pi^{3/2} \lambda^3_{e}  {\bf P}^{k^\prime} \int_C \frac{dz}{2\pi i} e^{-\beta z} \sum_{n=1}^\infty \sum_{l=0}^{n-1} \sum_{m=-l}^{l} \frac{1}{\left(E_n-z\right)^2} \left< nlm \left| \frac{m\omega^2_c}{8} r^2 sin^2 \theta   \right| nlm \right> \, .
\end{equation}
The calculation of the matrix elements is readily carried out \cite{Landau}, with the result
\begin{eqnarray} 
\left< nlm \left| \frac{m_e\omega^2_c}{8} r^2 sin^2 \theta   \right| nlm \right>  & =  & \frac{m_e \omega^2_c}{8} a^2_B \frac{n^2}{2} \left(5 n^2+1-3 l (l-1) \right) \, \nonumber\\
& \times & \frac{4l^3+6l^2+2(2l+1)m^2-2l-2}{(2l+1) (2l-1) (2l+3)}  \, . 
\end{eqnarray}
With that we obtain after integration and summation over the magnetic quantum number m
\begin{equation} 
\label{209} B^{2b}_{ei}  = -4 \pi^{3/2} \lambda^3_{e}  \frac{x_e^2}{3 \xi_{e}^2} {\bf P}^{k^\prime} \sum_{n=1}^\infty \sum_{l=0}^{n-1} e^{\left(\frac{1}{n^2} \frac{\xi_{e}}{2} \right)^2 } n^2 \left(5 n^2+1-3 l (l+1) \right) \, \frac{8l^3+12l^2-2l-3}{(2l-1) (2l+3)} \, .
\end{equation}
By summing over l we get
\begin{equation} 
\label{210} B^{2b}_{ei}  = -4 \pi^{3/2} \lambda^3_{e} {\bf P}^{k^\prime} \sum_{n=1}^\infty e^{-\beta E_n} \frac{x_e^2}{6 \xi^2_{e}} n^4 \left(7 n^2+5 \right) \, .
\end{equation}
As before we expand the exponential, introduce the $\zeta$-function and obtain the following expression for the bound state contribution 
\begin{equation}
\label{211} B^{2b}_{ei} =-4 \pi^{3/2} \lambda^3_{e} \frac{x_e^2}{24}  \sum_{k=6} \frac{7 \zeta(k-4)+5 \zeta(k-2)}{\Gamma(2+k/2)} \left(\frac{\xi_{e}}{2}\right)^k \, .
\end{equation}
Now we shall calculate the scattering part. This contribution may be obtained by applying the same arguments that have led to the final expression of $B^1_{ab}$. 
\begin{equation}
\label{212} B^2_{ei} =-2 \pi^{3/2} \lambda^3_{e} \frac{x_e^2}{24}  \sum_{k=6} \frac{7 \zeta(k-4)+5 \zeta(k-2)}{\Gamma(2+k/2)} \left(\frac{\xi_{e}}{2}\right)^k \, .
\end{equation}
Again, only contributions $k \geq 6$ are retained from this sum.  

\subsubsection{Final results for the electron-ion contribution}
We can now take the sum of the various contributions Eqs.(\ref{2201}, \ref{207} and \ref{212}) in order to obtain the quantum virial function. As we have indicated before we may now drop the operator ${\bf P}^{k^\prime}$ and may postulate the virial coefficient, with 
\begin{equation}
\label{213} B^d_{ei}  =  2 \pi \lambda_{e}^3 Q^0 (\xi_e) +  2 \pi \lambda_{e}^3 \frac{x_e^2}{24} Q^B (\xi_{e})  \, ,
\end{equation}
where we have defined the new magnetic quantum virial function $Q_{ab}^B$ by
\begin{equation}
\label{3001} Q^B (\xi_{ab}) = \sqrt \pi \sum_{k=2}^{\infty} \frac{ (k-3) \zeta(k-2) + (k-5) \zeta(k-4) }{\Gamma(2+k/2)} \left(\frac{\xi_{ab}}{2}\right)^k \, .
\end{equation}
In spite of the fact that the derivation given above is valid only for $k \geq 6$ we have extended the sum to $k \geq 2$. By studying the asymptotic properties of this function we will show that the magnetic quantum virial function has the correct asymptotics for large $\xi$. Another independent verification of this result can be obtained by expanding the exact second order contribution as given in Eq.(\ref{mwt9}). The quantum virial function $Q^B (\xi_e)$ may be interpreted as the limit of $Q^B (\xi_{ei})$ with an infinite proton mass $m_i \rightarrow \infty$ . For $k=3,5$ we make use of the relation $\lim_{k\rightarrow 3} (k-3) \zeta(k-2) =1 $. \par
In the next section we will show that the same analytical function determines also the contribution of the electron-electron interaction. 
  
\subsection{Electron-electron contribution}
We first study the Hamilton operator in c.m. and relative coordinates. In this case the hamiltonian is separable and may be written as $H_{ee} = H_{ee}^{cm} + h_{ee}$, with the center of mass hamiltonian
\begin{equation}
H_{ee}^{cm} = \frac{{\bf P}^2}{4 m_e}+\frac{e^2 B^2}{4 m_e} R^2 \sin \Theta^2  -i \frac{\hbar \omega_c}{2} \frac{\partial}{\partial \Phi} \, .
\end{equation}
It describes the free motion of a particle with mass $2m_e$ parallel to the field. While we have an harmonic oscillator with frequency $\omega_c=eB/m_e$ perpendicular to the field. The hamiltonian for the relative motion is given by
\begin{equation}
h_{ee}= \frac{{\bf p}^2}{m_e}+\frac{e^2 B^2}{16 m_e} r^2 \sin \theta^2 -i \frac{\hbar \omega_c}{2} \frac{\partial}{\partial \phi}+ \frac{e^2}{4\pi \epsilon_0 r} \, .
\end{equation}
It has the same structure as the hamiltonian for the relative motion of an electron in the field of a proton with infinite mass. The only difference is the appearance of different masses in the various terms of $h_{ei}$ and $h_{ee}$. However, by appropriately redefining the length scales and dimensionless parameters involved in the problem, one can map $h_{ee}$ onto $h_{ei}$. This means in detail the replacement of $\lambda_e$ by $\lambda_{ee}$ and of $\xi_e$ by $\xi_{ee}$ in Eq.(\ref{213},\ref{3001}). Now we may use the analyticity of the virial coefficient with respect to the interaction parameter. We may extend the result obtained for the electron-ion part Eq.(\ref{3001}) by analytical continuation to negative $\xi$-values. Thus we have for the electron-electron contribution
\begin{equation}
B^d_{ee} =  2 \pi \lambda_{ee}^3 Q^0(\xi_{ee})+ 2 \pi \lambda_{ee}^3 \frac{x_e^2}{24}   Q^B (\xi_{ee}) \, .
\end{equation}
The magnetic quantum virial function $Q^B (\xi_{ee})$ is given by Eq.(\ref{3001}). Note that this series holds also for the ion-ion interaction if $m_e$ is substituted by $m_i$. However its contribution to the virial coefficient is negligible in the weak field limit. \par

Let us briefly state the result for the exchange part of the electron-electron contribution. It may be obtained by introducing an additional factor $(-1)^l$ in Eq.(\ref{194} and \ref{208}) which takes into account the exact symmetry of the wavefunction. Then following the same steps as described in section III we find 
\begin{equation}
\label{215} B_{ee}^{ex} = - \pi \lambda_{ee}^3 \frac{\cosh(2x_e)}{\cosh^2(x_e)} E^0(\xi_{ee}) - \pi \lambda_{ee}^3 \frac{x_e^2}{6}   \frac{\cosh(2x_e)}{\cosh^2(x_e)} E^B (\xi_{ee})  \, ,
\end{equation}
with the new magnetic exchange virial function 
\begin{equation}
\label{216} E^B (\xi_{aa}) = \sqrt \pi \sum_{k=0}^{\infty} \frac{1}{\Gamma(1+k/2)} \left( \frac{k}{2+k} (1-2^{4-k}) \zeta(k-3) - \frac{4}{2+k} (1-2^{2-k}) \zeta(k-1)\right) \left(\frac{\xi_{aa}}{2}\right)^k \, .
\end{equation}
The factor $\cosh(2x_e)/\cosh^2(x_e)$ in Eq.(\ref{216}) is a result of the spins of the particles and can be calculated exactly. 
Again, one may check these results for the order $k=0,1$ by comparison with the exact contributions given by Eqs.(\ref{2001} and \ref{2002}).  
\subsection{Asymptotic properties of the virial function}
Let us now make an independent test of the above made statements. This investigation relies on two facts. First we consider the elctron-electron contribution only, then in the limit $\xi \rightarrow \infty $ the quantum virial function $Q^B (\xi)$ should be equal to the Wigner-Kirkwood expansion \cite{AlJa79}, since $\xi \sim \hbar^{-1} $. That means in this limit the plasma behaves essentially as a classical system. The second argument is that the electron-electron contribution may be obtained from the ion-electron contribution, and vice versa, by simple replacements of the interaction parameter as discussed in the previous section. Let us start by studying the higher order contributions $k \geq 6$ to the magnetic virial function (truncated virial function), which read according to Eq.(\ref{3001}) as 
\begin{equation}
\label{4001} Q^{\prime B} (\xi) = \sqrt \pi \sum_{k=6}^{\infty} \frac{ (k-3) \zeta(k-2) + (k-5) \zeta(k-4) }{\Gamma(2+k/2)} \left( \frac{\xi}{2}\right)^k \, ,
\end{equation}  
with $\xi<0$. The $\Gamma$-function can be represented by an inverse Laplace transform
\begin{equation}
\frac{1}{\Gamma(z)} = \int_{\delta-i\infty}^{\delta+i\infty} \frac{dt}{2\pi i} \frac{e^t}{t^z}  \, .
\end{equation}
With that and after rearranging the sum over k, $ Q^{\prime B} (\xi) $ can be rewritten as
\begin{eqnarray}
Q^{\prime B} (\xi) & = & \sqrt \pi \int_{\delta-i\infty}^{\delta+i\infty}  \frac{dt}{2\pi i} \frac{e^{t}}{t^2} \sum_{k=4}^{\infty} (k-3) \zeta(k-2) \left(1+ \left(\frac{\xi}{2 \sqrt{ t}}\right)^2\right) \left(\frac{\xi}{2 \sqrt{t}}\right)^k \nonumber\\
& - & \sqrt \pi \frac{\zeta(2)}{\Gamma(4)} \left(\frac{\xi}{2 }\right)^4- \sqrt \pi \frac{2 \zeta(3)}{\Gamma(9/2)} \left(\frac{\xi}{2 }\right)^5 \, .
\end{eqnarray} 
In the following we make use of the relation
\begin{equation}
\sum_{k=4}^\infty (-1)^k (k-3) \zeta(k-2) x^k = x^4 \left(\psi^\prime \left(x\right)-\frac{1}{x^2} \right) \, , \, \, \, x>0 \, , 
\end{equation}
which gives then
\begin{eqnarray}
Q^{\prime B} (\xi) & = & \sqrt \pi \int_{\delta-i\infty}^{\delta+i\infty}  \frac{dt}{2\pi i} \frac{e^{t}}{t^2} \left(\frac{\xi}{2 \sqrt{t}} \right)^4 \left(1+ \left(\frac{\xi}{2 \sqrt{t}}\right)^2 \right) \left(\psi^\prime \left(\frac{|\xi|}{2 \sqrt{t}}\right)-\left(\frac{2\sqrt{t} }{|\xi|}\right)^2\right) \nonumber\\
& - & \sqrt \pi \frac{\zeta(2)}{\Gamma(4)} \left(\frac{\xi}{2}\right)^4- \sqrt \pi \frac{2 \zeta(3)}{\Gamma(9/2)} \left(\frac{\xi}{2 }\right)^5 \, .
\end{eqnarray} 
It useful to employ the asymptotic expansion of the $\psi$ function
\begin{equation}
\psi(x)=\ln x -\frac{1}{2x}-\sum_{s=1}^m \frac{B_{2s}}{2s \, x^{2s}} + r_m(x) \, ,
\end{equation}
with the Bernoulli numbers $B_{2k}$. Then we can perform the inverse Laplace transform and find the following asymptotic expansion of the truncated magnetic quantum virial function 
\begin{eqnarray}
\label{4100} Q^{\prime B} (\xi) & = & - \frac{\sqrt \pi}{\Gamma(9/2)} \left( 1 + 2 \zeta(3) \right) \left(\frac{\xi}{2}\right)^5- \frac{\sqrt \pi}{\Gamma(4)} \left( \frac{1}{2} + \zeta(2) \right) \left(\frac{\xi}{2}\right)^4-\frac{\sqrt \pi}{\Gamma(7/2)} \left( 1 + B_2 \right) \left(\frac{\xi}{2}\right)^3 \nonumber\\
\label{4101} & - & \frac{\sqrt \pi}{2 \Gamma(3)} \left(\frac{\xi}{2}\right)^2 - \frac{\sqrt \pi}{\Gamma(5/2)} \left( B_2 + B_4 \right) \left(\frac{\xi}{2}\right)- \frac{\sqrt \pi}{\Gamma(3/2)}(B_4+ B_6) \left(\frac{2}{\xi}\right)- \frac{\sqrt \pi}{\Gamma(1/2)} (B_6+B_8) \left(\frac{2}{\xi}\right)^3+ o(\xi^{-5}) \, .
\end{eqnarray}
Now we may conclude that the full magnetic virial function defined by
\begin{equation}
Q^{B} (\xi) = \sqrt \pi \sum_{k=2}^{5} \frac{ (k-3) \zeta(k-2) + (k-5) \zeta(k-4) }{\Gamma(2+k/2)} \left(\frac{\xi}{2}\right)^k +Q^{\prime B} (\xi) 
\end{equation}
has the following asymptotic representation
\begin{equation}
Q^{B} (\xi) = - \frac{4}{45} \xi + \frac{4}{105}\frac{1}{\xi}+\frac{8}{105} \frac{1}{\xi^3} + o(\xi^{-5})\, .
\end{equation}
Remarkably, this procedure is accompanied by a term by term cancellation of the lower order contributions $k<6$ coming from the Taylor expansion with those coming from the asymptotic expansion. The final expression may now be compared with the $\hbar^2$ expansion , which was computed by Alastuey and Jancovici \cite{AlJa79}. 
The linear term in the asymptotic expansion of $ Q^B (\xi) $  is the term proportional to $\hbar^2$-term of the Wigner Kirkwood expansion and coincides with that of Alastuey and Jancovici. In addition to that we have found higher order contributions proportional to $\hbar^4$ and $\hbar^6$. With this derivation we have shown that that the magnetic virial function (\ref{3001}) has the correct asymptotic properties. This may be regarded as a strong support of the argument that the sum in Eq.(\ref{4001}) can be extended to the values of k=2,3,4, and 5, in order to obtain the desired result as given in Eq.(\ref{3001}). 

Notice that from the Wigner-Kirkwood expansion follows the absence of the linear term in the Taylor expansion (\ref{3001}).

Finally we give the asymptotic form of $Q^{B} (\xi)$  for positive arguments. To establish this property, we first observe that the magnetic virial function obeys the following relation
\begin{equation}
Q^{B} (\xi)+Q^{B} (- \xi) = 2 \sqrt \pi \sum_{k=2,4,\cdots}^{\infty} \frac{ (k-3) \zeta(k-2) + (k-5) \zeta(k-4) }{\Gamma(2+k/2)} \left(\frac{\xi}{2}\right)^k \, .
\end{equation}
From this, it follows by using the representation of the $\zeta$-function as an infinite sum and then carrying out the sum over k, that  
\begin{equation}
\label{5101} Q^{B} (- \xi) = Q^{B} (\xi) + \frac{\sqrt \pi}{8} \xi^2  +\frac{\sqrt \pi}{96} \left(\frac{\pi^2}{3} +1 \right) \xi^4 +2 \sqrt \pi \sigma^B (\xi) \, ,
\end{equation}
where we have defined 
\begin{eqnarray}
\label{5102} \sigma^B (\xi) & =  & \sum_{n=1}^\infty 2 n^2 \left(1+n^2\right) \left[e^{\left(\frac{\xi}{2}\right)^2 \frac{1}{n^2}} -1- \left(\frac{\xi}{2}\right)^2 \frac{1}{n^2}- \frac{1}{2 !} \left(\frac{\xi}{2}\right)^4 \frac{1}{n^4} \right] \nonumber\\
 & - &  \sum_{n=1}^\infty  n^4 \left(5+7 n^2\right) \left(\frac{2}{\xi}\right)^2 \left[e^{\left(\frac{\xi}{2}\right)^2 \frac{1}{n^2}} -1- \left(\frac{\xi}{2}\right)^2 \frac{1}{n^2}-  \frac{1}{2 !} \left(\frac{\xi}{2}\right)^4 \frac{1}{n^4} -  \frac{1}{3 !} \left(\frac{\xi}{2}\right)^6 \frac{1}{n^6} \right] \, .
\end{eqnarray}
Now let us briefly summarize the properties of the magnetic quantum virial function. In Fig.1 we have plotted $Q^{B} (\xi)$ for both positive and negative arguments, i.e. for electron-ion and electron-electron interaction , respectively. It shows an asymmetric behavior. For opposite charged particles the magnetic quantum virial function increases exponentially at large $\xi$, i.e. at low temperatures, due to the formation of bound states. While for like charged $Q^{B} (\xi)$ increases linear at large $\xi$. \\
The behavior of the exchange magnetic virial function is shown in Fig.2. In the quantum regime, at small $\xi$, one finds a finite contribution to the thermodynamic funcitons. While $E^{B} (\xi)$ decreases exponentially in the classical regime, i.e. at large $\xi$-values. This result was also found in \cite{AlJa79}. 

\section{Magnetization and magnetic susceptibility}
We now compute the magnetization in linear approximation (weak-field limit) and construct from this the magnetic susceptibility. Thereby spin effects and orbital effects are treated on equal footing. Let us suppose that the magnetization may be divided into ideal and interaction contributions
\begin{equation}
M = M_{id} + M_{int} \, .
\end{equation}
We restrict ourselves to the magnetization of the electronic subsystem, since the magnetization of the subsystem of the heavy positive ions is negligible small. However the contribution of the electron-ion interaction is fully included into our calculation. The ideal magnetization $M= -(1/\Omega) (\partial F/ \partial B)$ may be calculated from
\begin{equation}
M_{id} = nkT \frac{\partial}{\partial B} \ln \left(\frac{n\Lambda_e^3}{2} \frac{x_e}{\tanh(x_e)} \right) \, .
\end{equation}
Evaluating this in the weak field limit, we get Landau's result for the sum of the spin magnetism and the diamagnetism, which reads
\begin{equation}
M_{L} = \frac{1}{6} \frac{n \hbar^2 e^2 \beta B}{m_e^2} \, .
\end{equation}
The interaction part of the magnetization may be expressed in terms of the magnetic virial function. By taking the derivative of the full second virial coefficient with respect to the magnetic field one obtains 
\begin{equation}
\label{M1} M_{int} =  \frac{1}{6} \frac{n \hbar^2 e^2 \beta B}{m_e^2} \left(\frac{\pi}{4} n\lambda^3_e Q^B_e+\frac{\pi}{4} n\lambda^3_{ee} Q^B_{ee} - 3 \pi n\lambda^3_{ee} E^0_{ee} - \frac{\pi}{2} n \lambda^3_{ee} E^B_{ee} \right) \, .
\end{equation}
Here we have introduced a density expansion of the thermodynamic functions, that can be obtained from the fugacity expansion by an iteration procedure, as discussed in \cite{BaEb71,EKK}. This expression may now be used to calculate the zero field magnetic susceptibility ($\chi  =  \left(\partial (M_L+M_{int})/\partial B \right)_{B=0}$), with the result
\begin{equation}
\label{M2} \chi =  \chi_L \left(1+ \frac{\pi}{4} n\lambda^3_e Q^B(\xi_e) + \frac{\pi}{4} n\lambda^3_{ee} Q^B(\xi_{ee}) - 3 \pi n\lambda^3_{ee} E^0(\xi_{ee}) -\frac{\pi}{2} n\lambda^3_{ee} E^B(\xi_{ee}) \right) \, .
\end{equation}
The first term is Landau's result for the magnetic susceptibility, $\chi_L= (1/6) (n \hbar^2 e^2 \beta/m_e^2) $, of an ideal system in Boltzmann statistics, while the next terms describe the density effects. These effects contain the interaction of the particles as well as the deviation from the Boltzmann statistics. Fig.3 shows the magnetic susceptibility as a function of the density parameter $n\lambda_{ee}^3$ of the system for various temperatures. In Fig.4 we have plotted the magnetic susceptibility as a function of the inverse temperature for various fixed densities. We find for $\xi_{ee}<1.2$ a decrease and for $\xi_{ee}>1.2$ an increase of the paramagnetic susceptibility. The transition from negative to positive corrections occurs at $T\sim 2 \times 10^5 K$. This non-monontonic dependence on the temperature is the result of two competiting effects. The first effect can be explained on the basis of an ideal quantum plasma. The exchange contribution of the order $n^2$, which describes the first deviation from the Boltzmann statistics, decreases the magnetic susceptibility. On the other hand, the interaction between the particles tends to increase the magnetic susceptibility. This effect becomes dominant at low temperatures, while at high temperatures the exchange effects are dominant. 

We note that for $\xi \gg 1$, i.e. for $T\ll 2 \times 10^5 K$, the contribution from the positive interaction parameter ($\xi>0$) may become very large due to its exponentiell increase with $1/T$. The region where a considerable number of bound states are formed, requires a special treatment \cite{EKK}. Clearly, this theory is restricted to the region in which $\mid \chi-\chi_{L}\mid /\chi_{L} <1$ is valid. \par
Finally, we mention that the magnetization and magnetic susceptibility of an OCP can be derived from the results of the TCP (\ref{M1},\ref{M2}). This limit is obtained by sending the mass of one species to infinity and the charge to zero while ensuring charge neutrality of the system. Then the magnetic susceptibility of an OCP in linear response reads
\begin{equation}
\label{M3} \chi^{OCP} =  \chi_L \left(1+ \frac{\pi}{4} n\lambda^3_{ee} Q^B(\xi_{ee}) - 3 \pi n\lambda^3_{ee} E^0(\xi_{ee}) -\frac{\pi}{2} n\lambda^3_{ee} E^B(\xi_{ee}) \right) \, .
\end{equation}
In the previous section, we have checked that this expression coincides with the Wigner-Kirkwood expansion derived by Alastuey and Jancovici \cite{AlJa80}.

\section[Acknowledgments]{Acknowledgments}
This work was supported by the Deutsche Forschungsgemeinschaft under grant\#Eb 126/5-1. We thank P.Martin for focussing our attention on this problem.
\\

\newpage

Figure Captions\\

Fig. 1 Plot of the magnetic quantum virial function $Q^B(\xi)$ . The positive branch ($\xi>0$) corresponds to the electron-proton interaction and the negative branch ($\xi<0$) to the electron-electron interaction.

Fig. 2 Plot of the exchange magnetic quantum virial function $E^B(\xi)$.

Fig. 3 Magnetic susceptibility as a function of the degeneracy parameter $n\lambda_{ee}^3$ for various temperatures (note that $| \xi | \sim \left(157000/T[K]\right)^{1/2}$).

Fig. 4 Magnetic susceptibility as a function of the of the coupling parameter $\Gamma$ (inverse temperature) for various fixed densities.

\newpage

\begin{figure}[h]
\centerline{\epsfig {figure=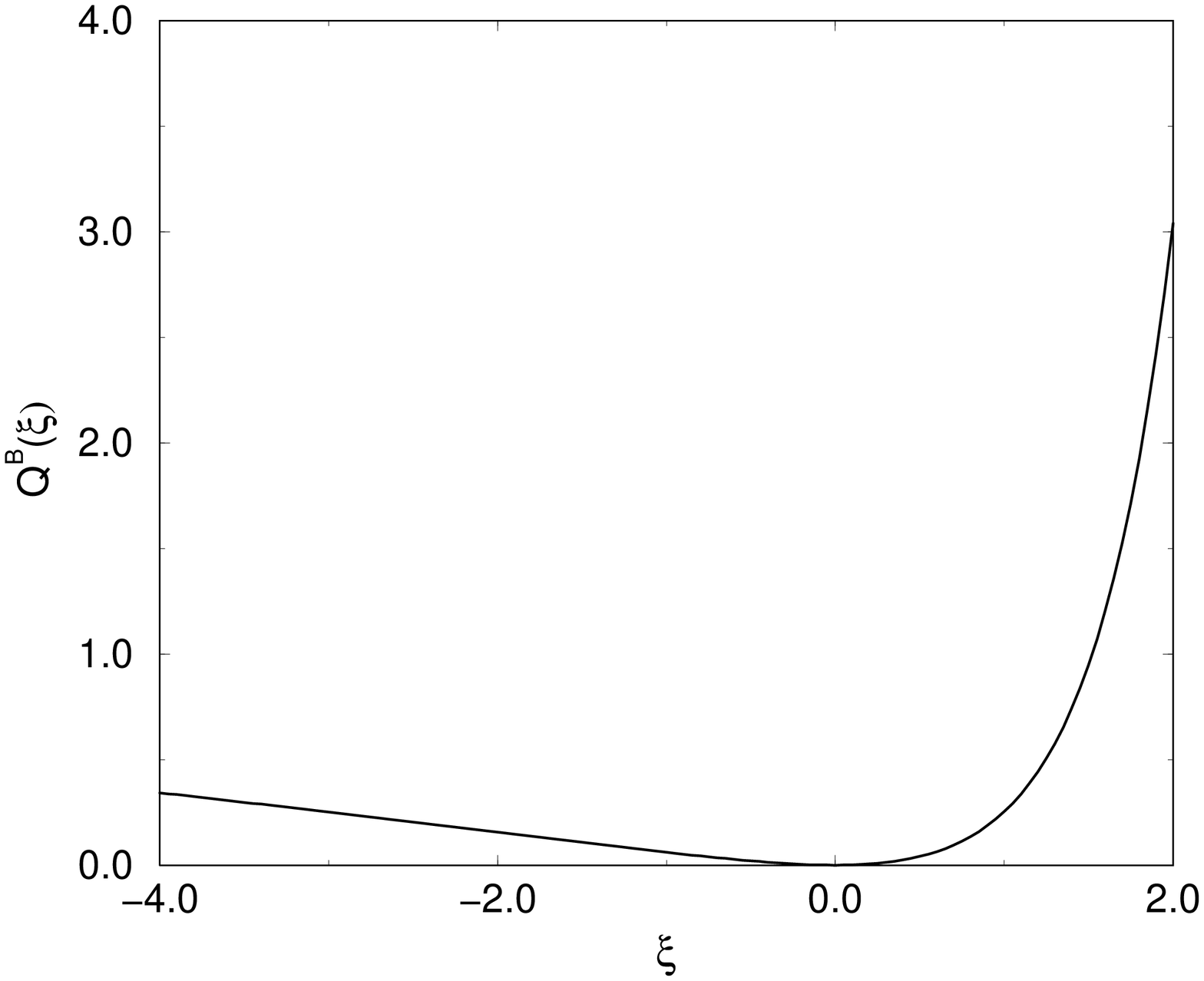,width=8.0cm,angle=-90}}
\end{figure}

\begin{figure}[h]
\centerline{\epsfig {figure=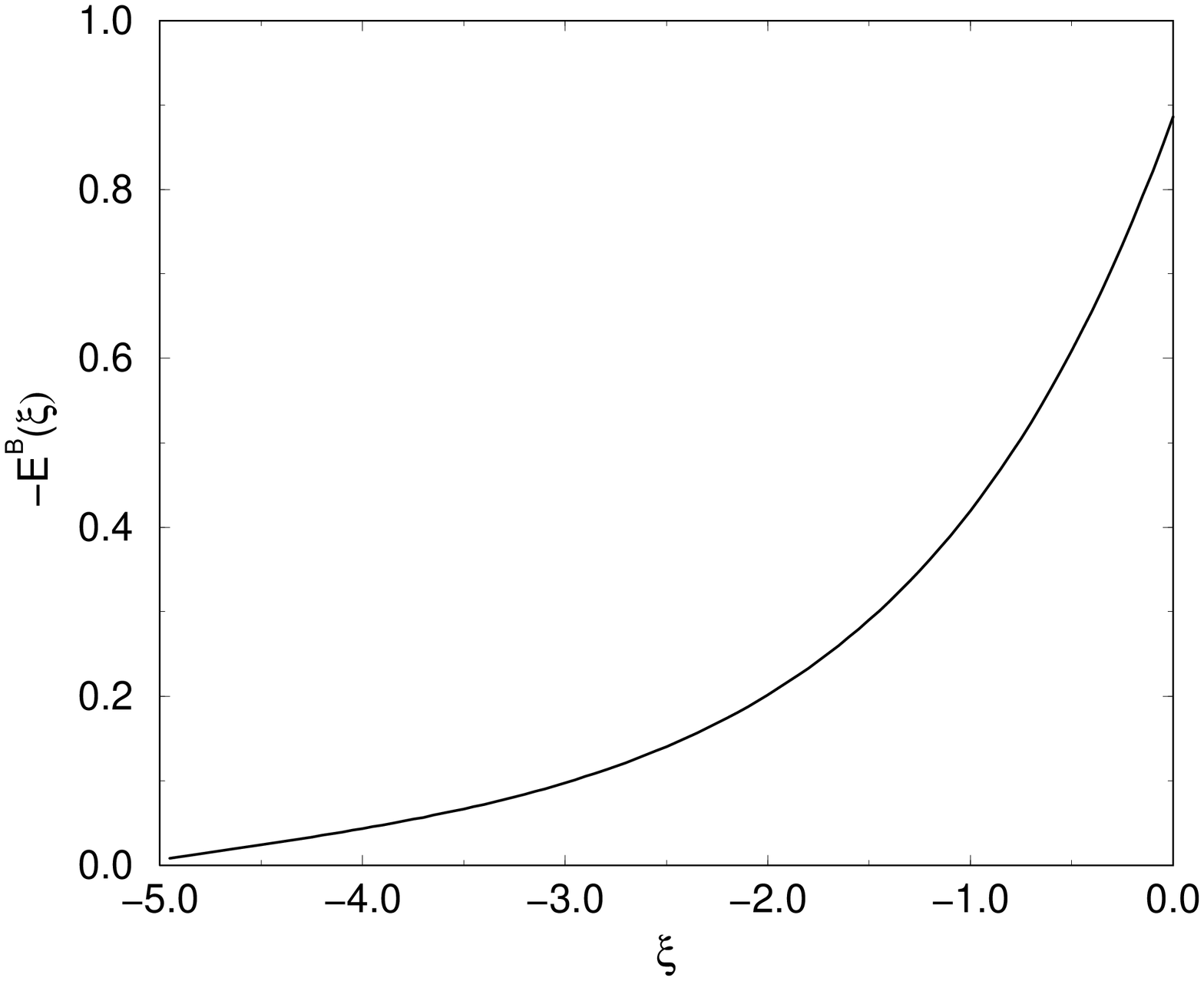,width=8.0cm,angle=-90}}
\end{figure}

\begin{figure}[h]
\centerline{\epsfig {figure=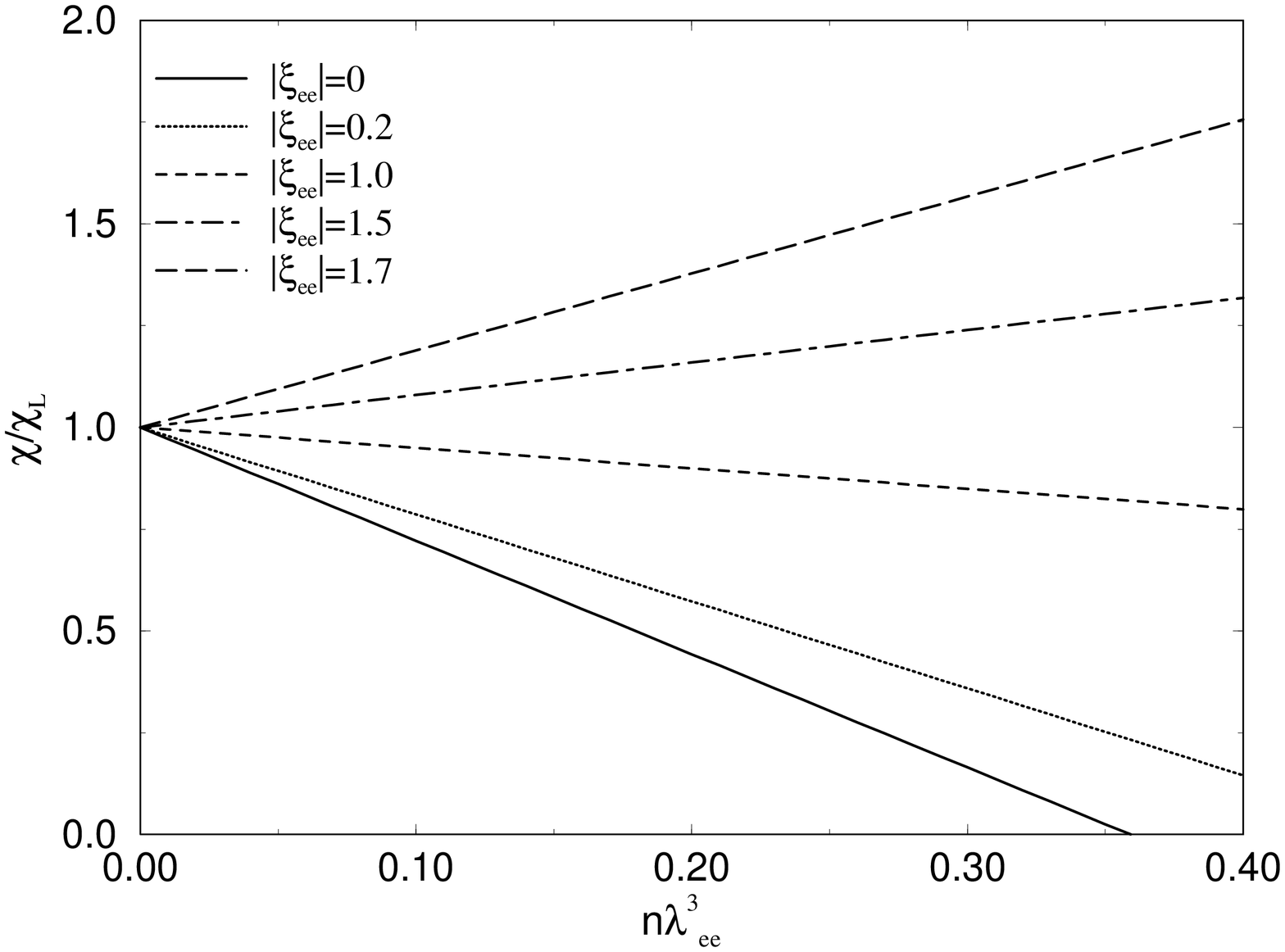,width=8.0cm,angle=-90}}
\end{figure}

\begin{figure}[h]
\centerline{\epsfig {figure=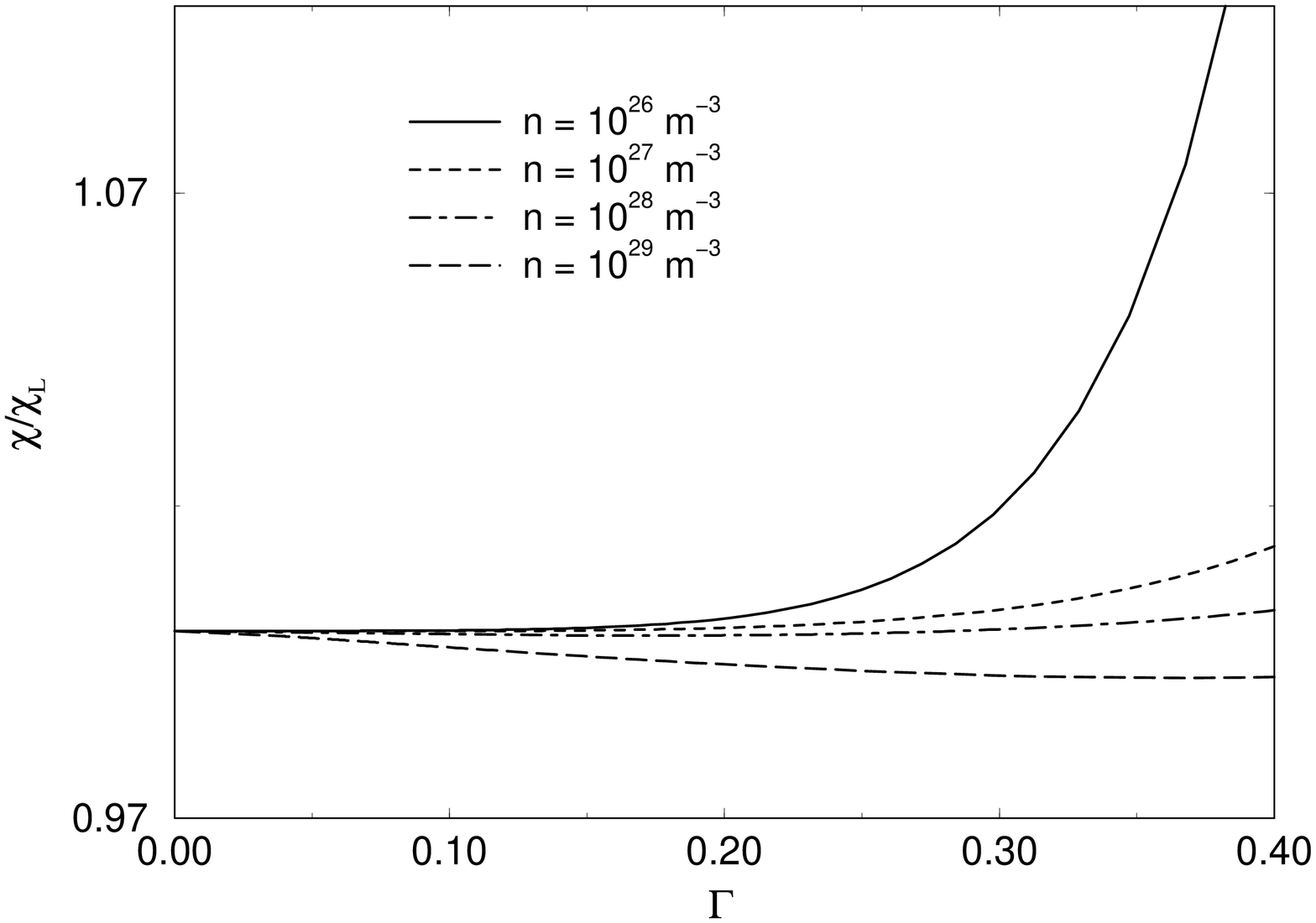,width=8.0cm,angle=-90}}
\end{figure}

\end{document}